\documentclass[11pt,preprintnumbers,aps,amssymb,nofootinbib,amsmath,superscriptaddress,notitlepage,prd]
{revtex4-2}
\usepackage{epsfig,epsf}
\usepackage{comment}
\usepackage{bm} 
\usepackage{bbm}
\usepackage{amsmath,amssymb,mathtools}
\usepackage{color} 
\usepackage{slashed} 
\usepackage{relsize}	
\usepackage{soul} 
\usepackage{hyperref}
\hypersetup{colorlinks=true, linkcolor=blue, citecolor=red, urlcolor=blue}
\usepackage{tensor} 
\usepackage{yfonts} 
\usepackage{orcidlink}
\newcommand{\beq}{\begin{equation}}
\newcommand{\beql}[1]{\begin{equation}\label{#1}}
\newcommand{\eeq}{\end{equation}}
\def\bal#1\gal{\begin{align}#1\end{align}}
\newcommand{\ball}[1]{\bal\label{#1}}
\newcommand{\eq}[1]{(\ref{#1})}
\newcommand{\fig}[1]{Fig.~\ref{#1}}
\renewcommand{\sec}[1]{Sec.~\ref{#1}}
\DeclareMathOperator{\im}{\mathrm{Im}}

\renewcommand{\b}[1]{{\bm #1}} 
\newcommand{\unit}[1]{\hat {{\bm #1}}} 

\begin{document}

\title{Rotational stability of magnetic field in rotating quark-gluon plasma}

\author{Aritra Das\orcidlink{0000-0002-3175-3473}}
\email{aritrad@iastate.edu, aritradasisu@gmail.com}

\author{Kirill Tuchin\orcidlink{0000-0002-6850-7698}}
\email{tuchink@gmail.com}

\affiliation{
Department of Physics and Astronomy, Iowa State University, Ames, Iowa, 50011, USA}

\date{\today}

\begin{abstract}
  
Relaxation of the magnetic field in rigidly rotating quark-gluon plasma is studied. It is shown that the infrared modes satisfying $k < |m|\Omega$ and $k< |m\pm 1|\Omega$, where integer $m$ is the projection of the orbital angular momentum along the rotating axis and $\Omega$ is the angular velocity, are unstable. The instability onset time and the magnetic field growth rate are computed for a standard initial profile of the magnetic field. Given the present phenomenological values of $\Omega$ and electrical conductivity $\sigma$ the instability is not expected to be a significant factor in the field's time evolution.

\end{abstract}

\maketitle

\section{Introduction}\label{sec:int}

Ultrarelativistic heavy-ion collisions produce quark-gluon plasma along with very strong magnetic field \cite{Kharzeev:2007jp,Skokov:2009qp,Voronyuk:2011jd,Ou:2011fm,Bzdak:2011yy,Bloczynski:2012en,Deng:2012pc}. The decaying field induces the eddy currents in plasma that significantly impact its time-evolution \cite{Tuchin:2010vs,Tuchin:2013apa,Tuchin:2013ie,Zakharov:2014dia,Tuchin:2015oka,Li:2016tel,Gursoy:2014aka,Gursoy:2018yai,Stewart:2017zsu}. It also possesses large vorticity, whose magnitude reaches $0.1$~fm$^{-1}$ and which points in the same direction as the magnetic field \cite{Csernai:2013bqa,Csernai:2014ywa,Becattini:2015ska,Deng:2016gyh,Jiang:2016woz,Kolomeitsev:2018svb,Deng:2020ygd,Xia:2018tes}. It is not unusual that rotation can render electromagnetic fields unstable \cite{velikhov1959stability,chandrasekhar1960stability,Balbus:1991ay,Wang:2023imu}. We therefore set about to investigate the rotational stability of the magnetic field produced in relativistic heavy-ion collisions.  

The basic set of equations describing the evolution of the magnetic field in plasma is 
\begin{align}
    \bm{\nabla}\times \bm{B} &=\bm{J}' +\b J+ \frac{\partial \bm{E}}{\partial t}, \label{eq:ME_curlB}\\
    \bm{\nabla}\cdot \bm{E} &= \varrho', \label{eq:ME_divE}\\
    \bm{\nabla}\cdot \bm{B} &= 0, \label{eq:ME_divB}\\
    \bm{\nabla}\times \bm{E} & = -\frac{\partial \bm{B}}{\partial t} \label{eq:ME_curlE}.
\end{align}
where $\b J'$ and $\varrho'$ are current and charge density of the valence quarks of the colliding ions and $\b J$ is the current density inside the plasma given by 
\bal\label{eq:J_total}
\b J = \sigma(\b E+\b v\times \b B)\,,
\gal
where $\sigma$ is electrical conductivity. The plasma motion is a whole consists of expansion and rotation. The expansion has a minor effect on the field evolution \cite{Stewart:2017zsu} and therefore will be neglected in our analysis. The rotation is described by $\b v= \b\Omega\times \b r$, where $\b \Omega$ is the angular velocity. We assume that $\b\Omega$ is constant vector parallel to the magnetic field. The consistent treatment of relativistically rotating systems  requires taking account of the causality constraint that imposes the boundary condition on the light-cylinder located at the radial distance $1/\Omega$ from the rotation axis. Since the relativistic heavy-ion phenomenology indicates that the light-cylinder of the quark-gluon plasma lies outside the plasma, its effect on the magnetic field can be neglected. 

The time-evolution of the magnetic field in plasma commences at the initial time $t=0$ when the value of the magnetic field and its time-derivative are given by $\b B_0$ and $\dot {\b B}_0$. These quantities can be obtained from the Maxwell equations describing the heavy-ions before their collision. The initial field is well approximated by the Gaussian profile that we adopt in this article. Another contribution to the magnetic, especially important at later times, comes about from the valence charges described by $\b J'$ and $\varrho'$. However, we  will ignore their contribution up until the last section where we present an argument why in fact it does not contribute much to the instability. 

The paper is organized as follows. In the next section we derive the differential equations satisfied by components of the magnetic field \sec{sec:master}. They are solved in  \sec{sec:modes}. \sec{sec:st} is dedicated to analysis of the dispersion relation which reveals the rotational instability. The stability criterion is derived. Our analysis is augmented in \sec{sec:numerics} by numerical solution of the magnetic field evolution with a particular initial condition relevant to the quark-gluon plasma.  The discussion and conclusions are presented in \sec{sec:summary}.


\section{Evolution equation of magnetic field}\label{sec:master}

We proceed to derive the equations that govern the time-evolution of  magnetic field. Taking the curl of  \eq{eq:ME_curlB} and using \eq{eq:J_total} and \eq{eq:ME_curlE}  we derive
\begin{align}\label{a1}
    \left(\frac{\partial^2}{\partial t^2}-\nabla^2 + \sigma \frac{\partial }{\partial t}\right) \bm{B}=  \sigma\bm{\nabla}\times(\bm{v}\times\bm{B})\,.
\end{align}
The last term on the right-hand-side of \eq{a1} can be transformed as 
\begin{align}\label{a3}
&\bm{\nabla}\times\left(\bm{v}\times\bm{B}\right)\vert_i =\left(\bm{\Omega}\times\bm{B}\right)_{i} - \bm{\Omega}\cdot(\bm{r}\times\bm{\nabla})B_i\,,
\end{align}
where we used $\b v= \b \Omega \times \b r$ with a constant angular velocity $\b \Omega$.
Therefore, the evolution equation of magnetic field reads
\begin{align}
    \left(\frac{\partial^2}{\partial t^2}-\nabla^2 + \sigma \frac{\partial }{\partial t}\right) \bm{B}=  \sigma\left[\left(\bm{\Omega}\times\bm{B}\right) - \bm{\Omega}\cdot(\bm{r}\times\bm{\nabla})\bm{B}\right]. \label{eq:B_evol_omega}
\end{align}
Assuming that  $\bm{\Omega} =\Omega\bm{\hat{z}}$ we obtain the following set of equations
\begin{align}
    &\left( \partial_{t}^2-\nabla^2+\sigma \partial_t \right)B_x =  \sigma \Omega \left[B_y-\left(\bm{r}\times\bm{\nabla}\right)_z B_x\right], \label{eq:B_x_evolution}\\
    &\left( \partial_{t}^2-\nabla^2+\sigma \partial_t \right)B_y =   \sigma \Omega \left[-B_x-\left(\bm{r}\times\bm{\nabla}\right)_z B_y\right], \label{eq:B_y_evolution}\\
    &\left( \partial_{t}^2-\nabla^2+\sigma \partial_t \right)B_z =  - \sigma \Omega\left(\bm{r}\times\bm{\nabla}\right)_z B_z. \label{eq:B_z_evolution}
\end{align}
Defining $\tilde{B}=B_x + i B_y$ and employing  Eqs.~\eq{eq:B_x_evolution} and \eq{eq:B_y_evolution} we derive  
\begin{align}
    \left[\partial_t^2 - \nabla^2 + \sigma \partial_t + \sigma\Omega(x\partial_y-y\partial_x) + i \sigma\Omega\right]\tilde{B} = 0\,.\label{eq:B_tilde_evolution_scaled}
\end{align}
Once $\tilde B$ is computed, its real and imaginary parts yield the transverse components $B_x$ and $B_y$. We can write \eq{eq:B_z_evolution} for the $z$-component in a similar form: 
\begin{align}
    \left[\partial_t^2 - \nabla^2 + \sigma \partial_t + \sigma\Omega(x\partial_y-y\partial_x)\right]B_z = 0. \label{eq:B_z_evolution_scaled}
\end{align}
The problem of solving Eqs~\eq{eq:B_tilde_evolution_scaled} and \eq{eq:B_z_evolution_scaled} consists in computing the eigenfunctions of the differential operator 
\bal\label{a5}
L= \partial_t^2 - \nabla^2 + \sigma \partial_t + \sigma\Omega(x\partial_y-y\partial_x)+\kappa i\sigma \Omega\,,
\gal
with $\kappa=0,\pm 1$. 

\section{The eigenmodes of magnetic field}\label{sec:modes}

To solve Eqs~\eq{eq:B_tilde_evolution_scaled} and \eq{eq:B_z_evolution_scaled} it is natural to employ cylindrical coordinates $(\rho,\phi,z)$: 
\begin{align}
    &\left(\partial_t^2-\partial_z^2-\partial_{\rho}^2-\frac{1}{\rho}\partial_{\rho}-\frac{1}{\rho^2}\partial_{\phi}^2+\sigma\partial_t + \sigma\Omega\partial_{\phi} + i\kappa\sigma\Omega\right)B(t,\b r) = 0\,,
    \label{eq:B_tilde_evolution_polar}
\end{align}
where $B$ stands for either $B_z$, $\tilde B$ or $\tilde B^*$ for $\kappa =0,  1,-1$ respectively. Separating variables we write 
\begin{align}
    B(t,\bm{r}) = \sum_{m=-\infty}^{\infty}\int\limits_{-\infty}^{\infty}\frac{dk_z}{2\pi}\int\limits_{0}^{\infty}dk_{\perp}\ k_{\perp}e^{i(m\phi +k_z z)}J_{m}(k_{\perp}\rho)\mathcal{B}_{m}(t,k_{\perp},k_z)\,, \label{eq:B_tilde_trial}
\end{align}    
where $\mathcal{B}_{m}(t,\b k)$ are amplitudes constrained by the initial conditions. Plugging \eq{eq:B_tilde_trial} into \eq{eq:B_tilde_evolution_polar} and employing the Bessel equation
\begin{align}\label{a7}
    \left(\partial_{\rho}^2+\frac{1}{\rho}\partial_{\rho}-\frac{m^2}{\rho^2}\right) J_{m}(k_{\perp}\rho) = -k_{\perp}^2 J_{m}(k_{\perp}\rho)\,
\end{align}
we obtain the equation that governs the time-evolution of the amplitudes:
\begin{align}\label{a9}
\partial_t^2\mathcal{B}_m(t,k_{\perp},k_z) + \sigma\,\partial_t \mathcal{B}_m(t,k_{\perp},k_z) + \left[k^2 + i\sigma(m+\kappa)\Omega\right] \mathcal{B}_m(t,k_{\perp},k_z) = 0,
\end{align}
where $|\bm{k}|^2 = k^2 = k_{\perp}^2 + k_z^2$. We are seeking a solution in the form $\mathcal{B}\sim e^{-i\omega t}$ which yields 
the dispersion relation:
\begin{align}\label{a11}
    \omega^2 + i\sigma \omega - k^2 - i\sigma (m+\kappa)\Omega = 0\,.
\end{align}
It admits two solutions
\begin{align}
    \omega^{\pm}_{m,\kappa}(k) = -i\frac{\sigma}{2}\pm \widetilde{\omega}_{m,\kappa}(k)= -i\frac{\sigma}{2}\pm \sqrt{k^2-\frac{\sigma^2}{4}+i\sigma (m+\kappa)\Omega}\,. \label{eq:modes}
\end{align}
The general solution to \eq{a9} is then
\ball{a13}
\mathcal{B}_{m,\kappa}(t,k_{\perp},k_z) = a^+_{m,\kappa}(k_{\perp},k_z)^{-i\omega^+_{m,\kappa}(k)t}+a^-_{m,\kappa}(k_{\perp},k_z)e^{-i\omega^-_{m,\kappa}(k)t}\,.
\gal
The coefficients $a^\pm_m(k_{\perp},k_z)$ are determined by the initial conditions at $t=0$:
\bal
a^+_{m,\kappa}(k_{\perp},k_z)= 
\frac{i\dot{\mathcal{B}}_{m,\kappa}(0,k_{\perp},k_z)-\omega^{-}_{m,\kappa}(k)\mathcal{B}_{m,\kappa}(0,k_{\perp},k_z)}{\omega^+_{m,\kappa}(k)-\omega^-_{m,\kappa}(k)}\,,\label{a15}\\
a^-_{m,\kappa}(k_{\perp},k_z)= -
\frac{i\dot{\mathcal{B}}_{m,\kappa}(0,k_{\perp},k_z)-\omega^{+}_{m,\kappa}(k)\mathcal{B}_{m,\kappa}(0,k_{\perp},k_z)}{\omega^+_{m,\kappa}(k)-\omega^-_{m,\kappa}(k)}\,.\label{a16}
\gal
The coefficients $\mathcal{B}_m(0,k_{\perp},k_z)$ and $\dot{\mathcal{B}}_m(0,k_{\perp},k_z)$ are extracted from the initial magnetic field and its derivative as 
\begin{align}
    \mathcal{B}_{m}(0,k_{\perp},k_z) &= \int\limits_{-\infty}^{\infty}dz\,e^{-ik_z z}\int\limits_{0}^{\infty}d\rho\  \rho\,J_m(k_{\perp}\rho)\int\limits_{0}^{2\pi}\frac{d\phi}{2\pi}e^{-im\phi}B(0,\b r) \label{eq:mathcalB_extraction}\\
    \dot{\mathcal{B}}_{m}(0,k_{\perp},k_z) &= \int\limits_{-\infty}^{\infty}dz\,e^{-ik_z z}\int\limits_{0}^{\infty}d\rho\  \rho\,J_m(k_{\perp}\rho)\int\limits_{0}^{2\pi}\frac{d\phi}{2\pi}e^{-im\phi}\dot{B}(0,\b r) \label{eq:dotmathcalB_extraction}
\end{align}
The expression of $\mathcal{B}_m(t,k_{\perp},k_z)$ can be simplified and written as  
\begin{align}\label{a18}
    &\mathcal{B}_m(t,k_{\perp},k_z) = \nonumber \\
    &e^{-\frac{\sigma}{2}t}\left\{\left[\cos\left[\widetilde{\omega}_{m,\kappa}(k) t\right] + \frac{\sigma}{2}\frac{\sin\left[\widetilde{\omega}_{m,\kappa}(k) t\right]}{\widetilde{\omega}_{m,\kappa}(k)}\right]\mathcal{B}_m(0,k_{\perp},k_z)+\frac{\sin\left[\widetilde{\omega}_{m,\kappa}(k) t\right]}{\widetilde{\omega}_{m,\kappa}(k)}\dot{\mathcal{B}}_m(0,k_{\perp},k_z)\right\}\,.
\end{align}

\section{The instability condition}\label{sec:st}

Information about the stability of the magnetic field in the rigidly rotating plasma is encoded in eigen-frequencies $\omega^{\pm}_m(k)$. In particular, the instability is indicated by the emergence of the positive imaginary part.
The imaginary part of the modes reads:
\begin{align}
   \im\omega_{m,\kappa}^{\pm}(k) = -\frac{\sigma}{2}\pm\frac{\text{sgn}(m+\kappa)}{\sqrt{2}}\left[\sqrt{\left(k^2-\frac{\sigma^2}{4}\right)^2+(m+\kappa)^2\sigma^2\Omega^2}-\left(k^2-\frac{\sigma^2}{4}\right)\right]^{1/2}, \label{eq:im_omega_pm}
\end{align}
where $\text{sgn}$ stands for the sign function. The instability of the field is signified by the singularities of the causal Green's function in the upper-half of the complex $\omega$-plane, which is indicated by the positive values of the imaginary part of the eigen-frequencies. Since the term inside the square root of the second term in \eqref{eq:im_omega_pm} is positive we are seeking a solution to the following inequalities:
\begin{align}
    \begin{cases}
        \im\omega^{+}_{m,\kappa}(k) > 0, & m+\kappa>0, \\
        \im\omega^{-}_{m,\kappa}(k) > 0, & m+\kappa<0\,.
    \end{cases}
\end{align}
These inequalities are satisfied by the modes with
\begin{align}
    k < |m+\kappa|\,\Omega\,.
    \label{eq:instability_condition}
\end{align}
It is remarkable that the instability condition \eqref{eq:instability_condition} does not depend on electrical conductivity $\sigma$. Nevertheless the logarithmic rate of the field increase $\partial_t\log B$, given by $\im\omega_{m,\kappa}^{\pm}$, clearly depends on it.  

\begin{figure}
    \centering
    \includegraphics[width=0.5\linewidth]{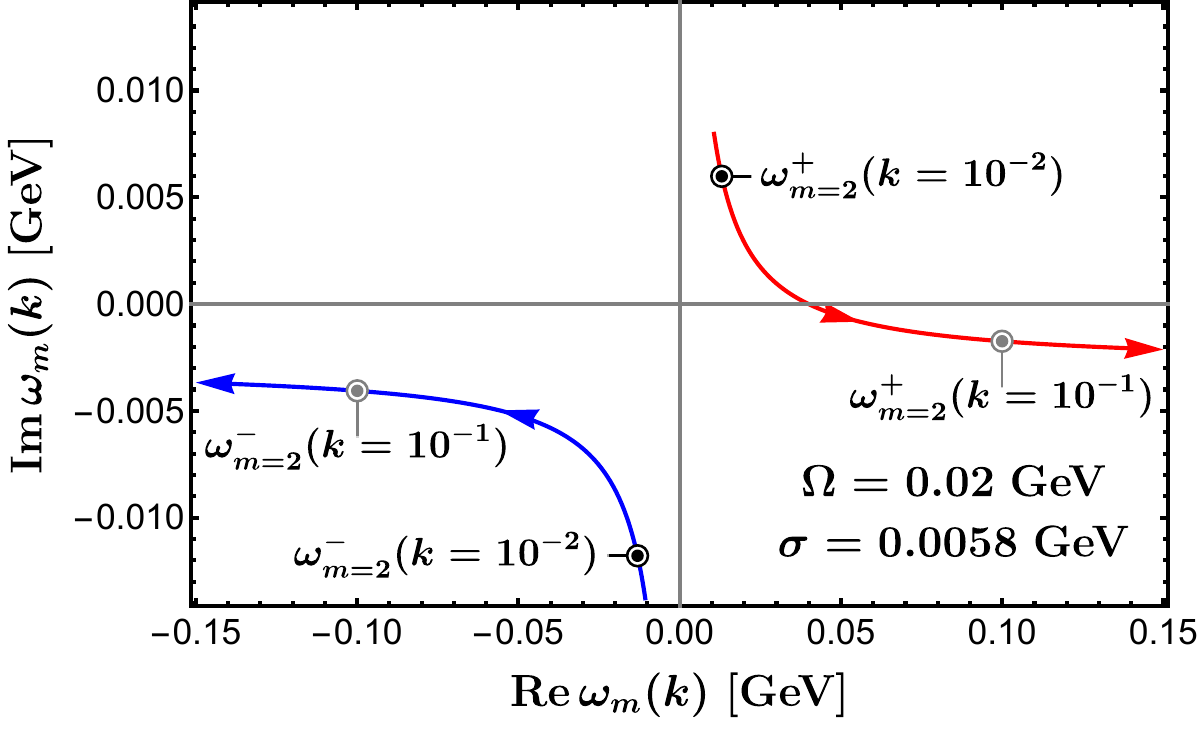}
    \caption{Eigen-frequencies of $B_z$, $\omega^{\pm}_{m}(k)$, in the complex $\omega$ plane. The instability is indicated by the positive imaginary part which vanishes when $k=m\Omega$. Arrows point in the direction of increasing $k$. } 
    \label{fig:hoping_of_poles}
\end{figure}

\begin{figure}
    \centering
    \includegraphics[width=0.45\linewidth]{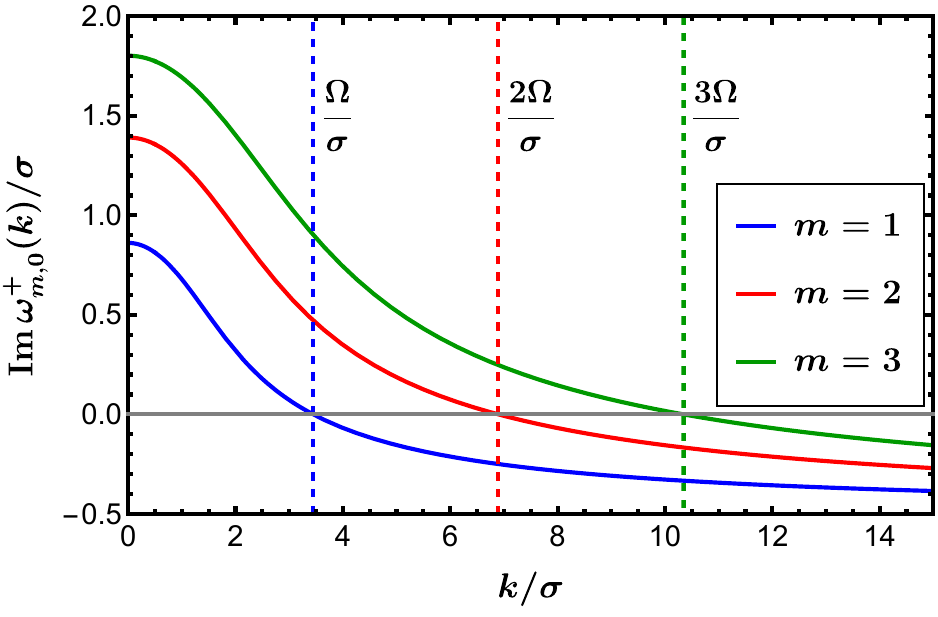}
    \includegraphics[width=0.45\linewidth]{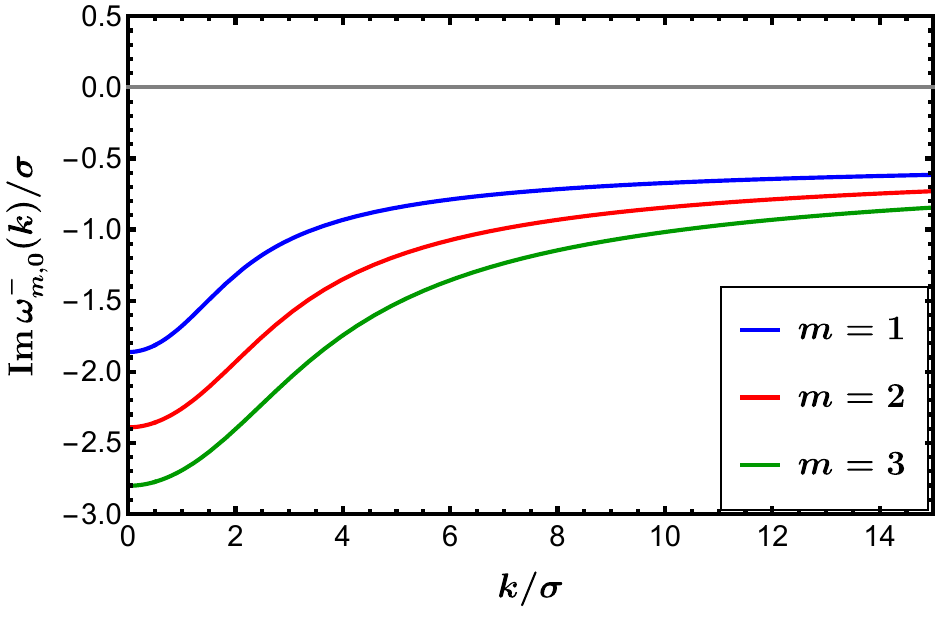}
    \caption{The figure shows the plot of $\text{Im}\,\omega_{m,0}$ with $k$ for different values of $m$. It demonstrates that for $k<m\Omega$, the $\omega^{+}_m(k)$ modes become unstable (left panel) whereas $\omega^{-}_m(k)$ modes remain stable for whole $k$ range (right panel). For negative values of $m$, the role of $\omega^{+}$ is exchanged with $\omega^{-}$.}
    \label{fig:modes_with_k}
\end{figure}

The eigen-frequencies $\omega^{\pm}_{m,\kappa}$ depend on two parameters $k$ and $m$ which run over all possible values constrained only by the initial conditions. Examination of Eq.~\eqref{eq:im_omega_pm} indicates that $\im \omega^{\pm}_{m,\kappa}$ is increasing function of $m$ at fixed $k$. 
At large $m$, it grows in proportion to $\sqrt{m}$. On the other hand, at fixed $m$ and large $k$, it behaves like
\begin{align}
    \im\,\omega_{m,\kappa}^{\pm}(k) \approx-\frac{\sigma}{2}\left(1 \mp \frac{(m+\kappa)\Omega}{k}\right)\,,\quad |m|\Omega\ll k\,.
\end{align}
Evidently, at large enough $k$ both modes cross into  the lower half-plane of complex $\omega$. Thus, the largest $\im\omega_{m,\kappa}^{\pm}$ is achieved at smaller values $k$ and larger values of $m$ as shown in Figs.~\ref{fig:hoping_of_poles} and \ref{fig:modes_with_k}.

\section{Evolution of magnetic field in rotating quark-gluon plasma}\label{sec:numerics}

In the previous section, we concluded that the magnetic field in rigidly rotating plasma is unstable. Whether this instability has a phenomenological significance hinges on the particular values of the parameters $\sigma$, $\Omega$ and the initial field profile. In this section, we investigate the time-evolution of the magnetic field in a system approximating the quark-gluon plasma. The initial magnetic field is taken to be 
\begin{align}\label{b1}
   \bm{B}(0,\bm{r}) = \lim_{t\to 0}\frac{B_0}{2}\left\{f(x,y-vt,z)+f(x,y+vt,z)\right\}\bm{\hat{z}}\,,
\end{align}
where 
\begin{align}\label{b3}
f(x,y,z)=\exp\left[ - \left(\frac{x^2}{\Delta_x^2}+\frac{y^2}{\Delta_y^2}+\frac{z^2}{\Delta_z^2}\right)\right]\,.
\end{align}
It is a superposition of two waves moving with velocity $v$ along the light-cones $y-vt$ and $y+vt$, where $y$ is the heavy-ion collision axis. In this model $\dot{\b B}(0,\b r)=0$. Since the average magnetic field points along the $z$-axis, we will focus only on computing the time-evolution of $B_z$. The instability of other components has a very similar form. Thus, throughout this section we set $\kappa=0$. The amplitudes $\mathcal{B}_m$ are obtained by inverting \eq{eq:B_tilde_trial} and using \eq{b1} which yields 
\begin{align}
    &\mathcal{B}_m(0,k_{\perp},k_z)\nonumber\\
    &= \int\limits_{-\infty}^{\infty} dz\, e^{-i k_z z} \int\limits_{0}^{\infty} d\rho\ \rho J_{m}(k_{\perp}\rho)\int\limits_{0}^{2\pi} \frac{d\phi}{2\pi}e^{-im\phi}
     B_0\unit z\exp\left[ - \left(\frac{\rho^2\cos^2\phi}{\Delta_x^2}+\frac{\rho^2\sin^2\phi}{\Delta_y^2}+\frac{z^2}{\Delta_z^2}\right)\right]\label{b3.1}\nonumber \\
    &=\begin{cases}
        B_0\unit z \frac{\sqrt{\pi}}{2}\Delta_x\Delta_y\Delta_z\exp\left(-\frac{1}{4}\left[\frac{k_{\perp}^2}{2}(\Delta_x^2+\Delta_y^2)+k_z^2\Delta_z^2\right]\right)I_{\frac{m}{2}}\left(\frac{k_{\perp}^2}{8}(\Delta_x^2-\Delta_y^2)\right), & \text{ even } m, \\
        0, & \text{ odd }m
    \end{cases} 
\end{align}
where $I$ stands for the modified Bessel function. The integrals are performed in the Appendix~\ref{sec:Coefficients}. From $\dot{\bm{B}}(0,\bm{r})=0$ and Eq.~\eqref{eq:dotmathcalB_extraction}, it is evident that $\mathcal{B}_m(0,k_{\perp},k_z) = 0$. Since $B(t,\bm{r})$ is real, the amplitudes $\mathcal{B}_m(t,k_{\perp},k_z)$ satisfy the equation:
\begin{align}
    \mathcal{B}^{*}_m(t,k_{\perp},k_z) = (-1)^m\mathcal{B}_{-m}(t,k_{\perp},-k_z),
\end{align}
which can be used to represent \eqref{eq:B_tilde_trial} as the sum of the two terms one of which is independent of $\Omega$:
\begin{align}
    &B(t,\bm{r}) = \nonumber \\
    &\int\limits_{-\infty}^{\infty}\frac{dk_z}{2\pi}\int\limits_{0}^{\infty}dk_{\perp}\ k_{\perp}J_{0}(k_{\perp}\rho) \cos(k_z z) \mathcal{B}_{0}(t,k_{\perp},k_z)\nonumber\\
    &+ 2\sum_{m=1}^{\infty}\int\limits_{-\infty}^{\infty}\frac{dk_z}{2\pi}\int\limits_{0}^{\infty}dk_{\perp}\ k_{\perp}J_{m}(k_{\perp}\rho)\,\text{Re}[e^{i(m\phi +k_z z)}\mathcal{B}_{m}(t,k_{\perp},k_z)]\,.
\end{align}
The fact that $\mathcal{B}_{m}(t,\b k)$ is even in $k_z$ for all $m$, allows us to write 
\begin{align}
    &B(t,\bm{r}) = \nonumber \\
    &\int\limits_{-\infty}^{\infty}\frac{dk_z}{2\pi}\int\limits_{0}^{\infty}dk_{\perp}\ k_{\perp}J_{0}(k_{\perp}\rho) \cos(k_z z) \mathcal{B}_{0}(t,k_{\perp},k_z)\nonumber\\
    &+ 2 \sum_{m=1}^{\infty}\int\limits_{-\infty}^{\infty}\frac{dk_z}{2\pi}\int\limits_{0}^{\infty}dk_{\perp}\ k_{\perp}J_{m}(k_{\perp}\rho)\cos(k_zz)\,\text{Re}[e^{im\phi}\mathcal{B}_{m}(t,k_{\perp},k_z)]. \label{eq:BzFinal_v1}
\end{align}
Substituting $\mathcal{B}_m(0,k_{\perp},k_z)$ described by \eqref{b3.1} and $\dot{\mathcal{B}}_m(0,k_{\perp},k_z) = 0$ into \eqref{a18} we obtain the amplitudes at time $t$:
\begin{align}
    &\mathcal{B}_m(t,k_{\perp},k_z) = B_0\frac{\sqrt{\pi}}{2}\Delta_x\Delta_y\Delta_z e^{-\frac{\sigma}{2}t}\Bigg[\cos\left[\widetilde{\omega}_{m}(k) t\right] \nonumber \\
    &\hspace{2cm} +\frac{\sigma}{2}\frac{\sin\left[\widetilde{\omega}_{m}(k) t\right]}{\widetilde{\omega}_{m}(k)}\Bigg]\exp\left(-\frac{1}{4}\left[\frac{k_{\perp}^2}{2}(\Delta_x^2+\Delta_y^2)+k_z^2\Delta_z^2\right]\right)I_{\frac{m}{2}}\left(\frac{k_{\perp}^2}{8}(\Delta_x^2-\Delta_y^2)\right)\, , \label{eq:BzFinal_v2}
\end{align}
where $\tilde{\omega}_{m}$ is given by \eqref{eq:modes} with  $\kappa=0$. The stationary limit $\Omega=0$, is discussed in detail in Appendix~\ref{sec:app:Omega=0}. 

\begin{figure}[ht]
\includegraphics[width=0.6\linewidth]{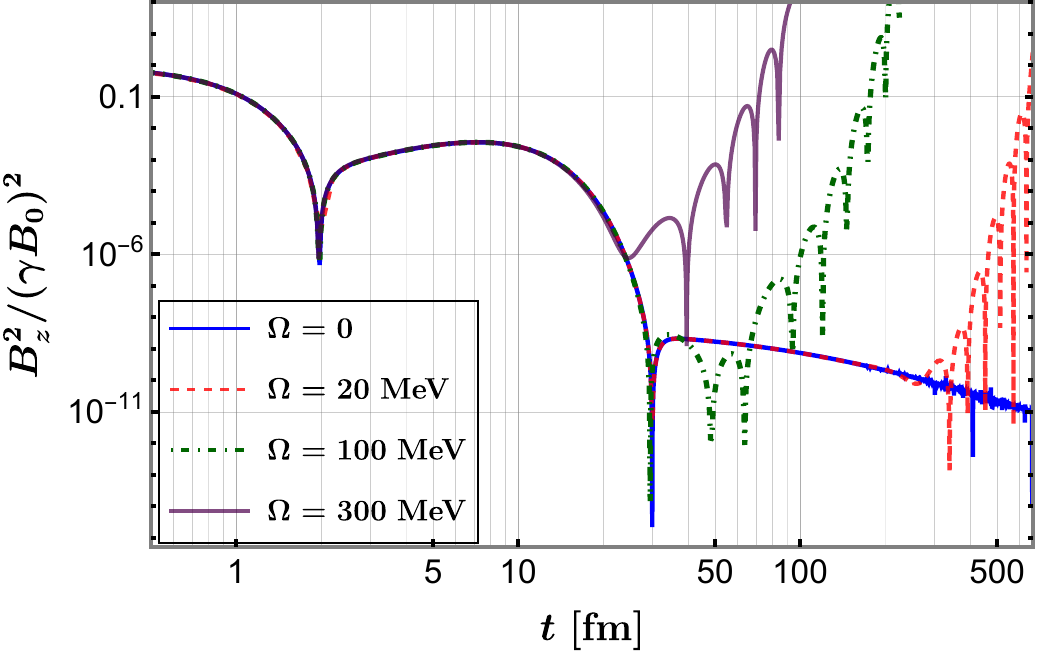}
  \caption{Time-dependence of magnetic field in heavy-ion collisions on the log-log scale at $(x,y,z) = (1,0,1)$~fm, Gaussian parameters in \eq{b3} are $\Delta_x = \Delta_z = 10$~fm and $\Delta_y = \Delta_x /\gamma$ with the Lorentz parameter $\gamma= (1-v^2)^{-1/2}=10$, electrical conductivity is $\sigma = 5.8$ MeV \cite{Aarts:2007wj,Ding:2010ga,Amato:2013oja}. We explore the model at late times even though the realistic plasma freezes out at about 10 fm/c.}
\label{fig:model_instability_fewOmega}
\end{figure}

\begin{figure}[ht]
    \centering
    \includegraphics[width=0.6\linewidth]{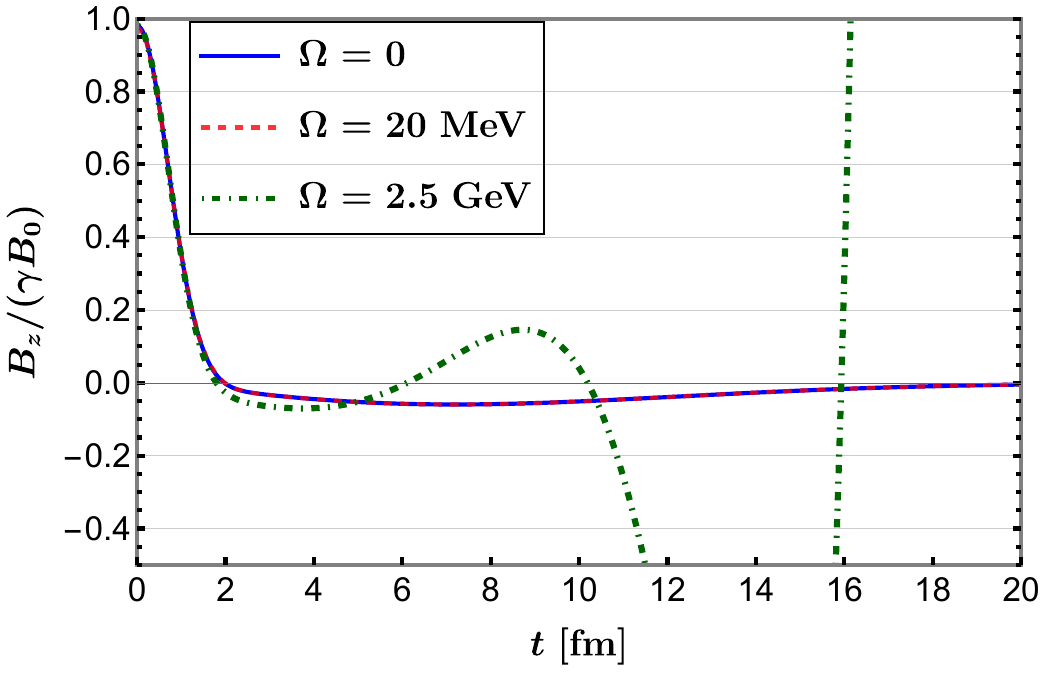}
    \caption{Time-dependence of magnetic field in heavy-ion collisions on the linear scale. The values of parameters, except $\Omega$, are the same as in \fig{fig:model_instability_fewOmega}. }
    \label{fig:model_high_Omega}
\end{figure}

The sum over $m$ and the integral over $k$ in \eqref{eq:BzFinal_v1} can be performed numerically to obtain the spacetime profile of $B_z$. Since the magnetic field instability is only mildly dependent on the coordinates, we fixed the observation point at $x = z = 1$~fm, $y=0$~fm. The results are exhibited in Figs.~\ref{fig:model_instability_fewOmega} and \ref{fig:model_high_Omega} for various angular velocities $\Omega$. The electrical conductivity is set at the value representative of the quark-gluon plasma. We plot $B_z^2$ on the log-log scale in \fig{fig:model_instability_fewOmega} and $B_z$ on the linear scale in \fig{fig:model_high_Omega}.  The realistic angular velocity does not exceed about $20$~MeV \cite{Csernai:2013bqa,Csernai:2014ywa,Becattini:2015ska,Deng:2016gyh,Jiang:2016woz,Xia:2018tes}. Nevertheless, in \fig{fig:model_instability_fewOmega} we studied much higher angular velocities in order to ascertain the typical time $\tau$ at which the instability sets in. We see that the instability is manifested earlier with increasing $\Omega$ in agreement with our previous discussions. More precisely, we can estimate 
\begin{align}\label{b5}
\tau (\text{fm})\approx \frac{20}{\Omega(\text{MeV})}\,. 
\end{align}
We note the magnetic field oscillations around zero. The sharp minima in the figure correspond to the vanishing magnetic field. The oscillatory behavior is expected to be less pronounced for specially averaged field.

One can wonder how fast should the rotation be in order that its effect be phenomenologically relevant for the quark-gluon plasma with lifetime about 10~fm/$c$. This is addressed in\fig{fig:model_high_Omega} we shows that $\Omega$ should be on the order of few GeV---two orders of magnitude above the values suggested by the current phenomenological models. At such values of $\Omega$ our calculation breaks down because it ignores the circumstance that the correlation length in plasma in the perpendicular plane $1\Omega$ is much shorter than the plasma size. The physics of such extremely rapidly rotating plasma is quite different from the non-rotating one \cite{Buzzegoli:2023yut}. 

\section{Discussion and Summary}\label{sec:summary}

We investigated the stability of magnetic field in uniformly rotating plasma and determined that soft modes with $k<|m+\kappa|\Omega$, where $\kappa=\pm 1,0$ are unstable regardless of the electric conductivity $\sigma$. However, the rate of growth, given by the imaginary part of frequency, does depend on $\sigma$.   It is interesting to note that the modes with $k<|m|\Omega$ are superradiant \cite{zel1971generation,zel1972amplification}. 

Our analysis assumed that rotation is non-relativistic, meaning that the $\Omega r\ll 1$. This allowed us to neglect the effects at the light-cylinder. This approximation holds well in practical applications. 

The contribution of the valence charge $q$ moving with velocity $\b v$ along the $y$-axis can be calculated by substituting $\b J'=  q\b v \delta(y-vt)\delta(\b b)$ into \eq{eq:ME_curlB}, where $\b b$ is the position vector in the $xz$ plane. Since the velocity is ultra-relativistic, $(\partial_t^2-\partial_y^2)B$ can be neglected compared to $(\partial_x^2+\partial_z^2)B$ \cite{Stewart:2017zsu,Tuchin:2015oka} which results in the dispersion relation that has no singularities in the upper-half plane of the complex $\omega$-plane. Thus, any contribution to the instability due to the valence charges is suppressed by at least one power of $1/\gamma^2$. 

In quark-gluon plasma produced in heavy-ion collisions, the instability develops only at unrealistically late times or high $\Omega$'s due to small electrical conductivity.

\acknowledgments
This work was supported in part by the U.S. Department of Energy Grants No.\ DE-SC0023692. We thank Nandagopal Vijaykumar and Siqi Xu for helping us run numerical computation.

\bibliography{EM_Fields}

\begin{thebibliography}{36}%
\makeatletter
\providecommand \@ifxundefined [1]{%
 \@ifx{#1\undefined}
}%
\providecommand \@ifnum [1]{%
 \ifnum #1\expandafter \@firstoftwo
 \else \expandafter \@secondoftwo
 \fi
}%
\providecommand \@ifx [1]{%
 \ifx #1\expandafter \@firstoftwo
 \else \expandafter \@secondoftwo
 \fi
}%
\providecommand \natexlab [1]{#1}%
\providecommand \enquote  [1]{``#1''}%
\providecommand \bibnamefont  [1]{#1}%
\providecommand \bibfnamefont [1]{#1}%
\providecommand \citenamefont [1]{#1}%
\providecommand \href@noop [0]{\@secondoftwo}%
\providecommand \href [0]{\begingroup \@sanitize@url \@href}%
\providecommand \@href[1]{\@@startlink{#1}\@@href}%
\providecommand \@@href[1]{\endgroup#1\@@endlink}%
\providecommand \@sanitize@url [0]{\catcode `\\12\catcode `\$12\catcode
  `\&12\catcode `\#12\catcode `\^12\catcode `\_12\catcode `\%12\relax}%
\providecommand \@@startlink[1]{}%
\providecommand \@@endlink[0]{}%
\providecommand \url  [0]{\begingroup\@sanitize@url \@url }%
\providecommand \@url [1]{\endgroup\@href {#1}{\urlprefix }}%
\providecommand \urlprefix  [0]{URL }%
\providecommand \Eprint [0]{\href }%
\providecommand \doibase [0]{https://doi.org/}%
\providecommand \selectlanguage [0]{\@gobble}%
\providecommand \bibinfo  [0]{\@secondoftwo}%
\providecommand \bibfield  [0]{\@secondoftwo}%
\providecommand \translation [1]{[#1]}%
\providecommand \BibitemOpen [0]{}%
\providecommand \bibitemStop [0]{}%
\providecommand \bibitemNoStop [0]{.\EOS\space}%
\providecommand \EOS [0]{\spacefactor3000\relax}%
\providecommand \BibitemShut  [1]{\csname bibitem#1\endcsname}%
\let\auto@bib@innerbib\@empty
\bibitem [{\citenamefont {Kharzeev}\ \emph {et~al.}(2008)\citenamefont
  {Kharzeev}, \citenamefont {McLerran},\ and\ \citenamefont
  {Warringa}}]{Kharzeev:2007jp}%
  \BibitemOpen
  \bibfield  {author} {\bibinfo {author} {\bibfnamefont {D.~E.}\ \bibnamefont
  {Kharzeev}}, \bibinfo {author} {\bibfnamefont {L.~D.}\ \bibnamefont
  {McLerran}},\ and\ \bibinfo {author} {\bibfnamefont {H.~J.}\ \bibnamefont
  {Warringa}},\ }\bibfield  {title} {\bibinfo {title} {{The Effects of
  topological charge change in heavy ion collisions: 'Event by event P and CP
  violation'}},\ }\href {https://doi.org/10.1016/j.nuclphysa.2008.02.298}
  {\bibfield  {journal} {\bibinfo  {journal} {Nucl. Phys. A}\ }\textbf
  {\bibinfo {volume} {803}},\ \bibinfo {pages} {227} (\bibinfo {year}
  {2008})},\ \Eprint {https://arxiv.org/abs/0711.0950} {arXiv:0711.0950
  [hep-ph]} \BibitemShut {NoStop}%
\bibitem [{\citenamefont {Skokov}\ \emph {et~al.}(2009)\citenamefont {Skokov},
  \citenamefont {Illarionov},\ and\ \citenamefont {Toneev}}]{Skokov:2009qp}%
  \BibitemOpen
  \bibfield  {author} {\bibinfo {author} {\bibfnamefont {V.}~\bibnamefont
  {Skokov}}, \bibinfo {author} {\bibfnamefont {A.~Y.}\ \bibnamefont
  {Illarionov}},\ and\ \bibinfo {author} {\bibfnamefont {V.}~\bibnamefont
  {Toneev}},\ }\bibfield  {title} {\bibinfo {title} {{Estimate of the magnetic
  field strength in heavy-ion collisions}},\ }\href
  {https://doi.org/10.1142/S0217751X09047570} {\bibfield  {journal} {\bibinfo
  {journal} {Int. J. Mod. Phys. A}\ }\textbf {\bibinfo {volume} {24}},\
  \bibinfo {pages} {5925} (\bibinfo {year} {2009})},\ \Eprint
  {https://arxiv.org/abs/0907.1396} {arXiv:0907.1396 [nucl-th]} \BibitemShut
  {NoStop}%
\bibitem [{\citenamefont {Voronyuk}\ \emph {et~al.}(2011)\citenamefont
  {Voronyuk}, \citenamefont {Toneev}, \citenamefont {Cassing}, \citenamefont
  {Bratkovskaya}, \citenamefont {Konchakovski},\ and\ \citenamefont
  {Voloshin}}]{Voronyuk:2011jd}%
  \BibitemOpen
  \bibfield  {author} {\bibinfo {author} {\bibfnamefont {V.}~\bibnamefont
  {Voronyuk}}, \bibinfo {author} {\bibfnamefont {V.~D.}\ \bibnamefont
  {Toneev}}, \bibinfo {author} {\bibfnamefont {W.}~\bibnamefont {Cassing}},
  \bibinfo {author} {\bibfnamefont {E.~L.}\ \bibnamefont {Bratkovskaya}},
  \bibinfo {author} {\bibfnamefont {V.~P.}\ \bibnamefont {Konchakovski}},\ and\
  \bibinfo {author} {\bibfnamefont {S.~A.}\ \bibnamefont {Voloshin}},\
  }\bibfield  {title} {\bibinfo {title} {{(Electro-)Magnetic field evolution in
  relativistic heavy-ion collisions}},\ }\href
  {https://doi.org/10.1103/PhysRevC.83.054911} {\bibfield  {journal} {\bibinfo
  {journal} {Phys. Rev. C}\ }\textbf {\bibinfo {volume} {83}},\ \bibinfo
  {pages} {054911} (\bibinfo {year} {2011})},\ \Eprint
  {https://arxiv.org/abs/1103.4239} {arXiv:1103.4239 [nucl-th]} \BibitemShut
  {NoStop}%
\bibitem [{\citenamefont {Ou}\ and\ \citenamefont {Li}(2011)}]{Ou:2011fm}%
  \BibitemOpen
  \bibfield  {author} {\bibinfo {author} {\bibfnamefont {L.}~\bibnamefont
  {Ou}}\ and\ \bibinfo {author} {\bibfnamefont {B.-A.}\ \bibnamefont {Li}},\
  }\bibfield  {title} {\bibinfo {title} {{Magnetic effects in heavy-ion
  collisions at intermediate energies}},\ }\href
  {https://doi.org/10.1103/PhysRevC.84.064605} {\bibfield  {journal} {\bibinfo
  {journal} {Phys. Rev. C}\ }\textbf {\bibinfo {volume} {84}},\ \bibinfo
  {pages} {064605} (\bibinfo {year} {2011})},\ \Eprint
  {https://arxiv.org/abs/1107.3192} {arXiv:1107.3192 [nucl-th]} \BibitemShut
  {NoStop}%
\bibitem [{\citenamefont {Bzdak}\ and\ \citenamefont
  {Skokov}(2012)}]{Bzdak:2011yy}%
  \BibitemOpen
  \bibfield  {author} {\bibinfo {author} {\bibfnamefont {A.}~\bibnamefont
  {Bzdak}}\ and\ \bibinfo {author} {\bibfnamefont {V.}~\bibnamefont {Skokov}},\
  }\bibfield  {title} {\bibinfo {title} {{Event-by-event fluctuations of
  magnetic and electric fields in heavy ion collisions}},\ }\href
  {https://doi.org/10.1016/j.physletb.2012.02.065} {\bibfield  {journal}
  {\bibinfo  {journal} {Phys. Lett. B}\ }\textbf {\bibinfo {volume} {710}},\
  \bibinfo {pages} {171} (\bibinfo {year} {2012})},\ \Eprint
  {https://arxiv.org/abs/1111.1949} {arXiv:1111.1949 [hep-ph]} \BibitemShut
  {NoStop}%
\bibitem [{\citenamefont {Bloczynski}\ \emph {et~al.}(2013)\citenamefont
  {Bloczynski}, \citenamefont {Huang}, \citenamefont {Zhang},\ and\
  \citenamefont {Liao}}]{Bloczynski:2012en}%
  \BibitemOpen
  \bibfield  {author} {\bibinfo {author} {\bibfnamefont {J.}~\bibnamefont
  {Bloczynski}}, \bibinfo {author} {\bibfnamefont {X.-G.}\ \bibnamefont
  {Huang}}, \bibinfo {author} {\bibfnamefont {X.}~\bibnamefont {Zhang}},\ and\
  \bibinfo {author} {\bibfnamefont {J.}~\bibnamefont {Liao}},\ }\bibfield
  {title} {\bibinfo {title} {{Azimuthally fluctuating magnetic field and its
  impacts on observables in heavy-ion collisions}},\ }\href
  {https://doi.org/10.1016/j.physletb.2012.12.030} {\bibfield  {journal}
  {\bibinfo  {journal} {Phys. Lett. B}\ }\textbf {\bibinfo {volume} {718}},\
  \bibinfo {pages} {1529} (\bibinfo {year} {2013})},\ \Eprint
  {https://arxiv.org/abs/1209.6594} {arXiv:1209.6594 [nucl-th]} \BibitemShut
  {NoStop}%
\bibitem [{\citenamefont {Deng}\ and\ \citenamefont
  {Huang}(2012)}]{Deng:2012pc}%
  \BibitemOpen
  \bibfield  {author} {\bibinfo {author} {\bibfnamefont {W.-T.}\ \bibnamefont
  {Deng}}\ and\ \bibinfo {author} {\bibfnamefont {X.-G.}\ \bibnamefont
  {Huang}},\ }\bibfield  {title} {\bibinfo {title} {{Event-by-event generation
  of electromagnetic fields in heavy-ion collisions}},\ }\href
  {https://doi.org/10.1103/PhysRevC.85.044907} {\bibfield  {journal} {\bibinfo
  {journal} {Phys. Rev. C}\ }\textbf {\bibinfo {volume} {85}},\ \bibinfo
  {pages} {044907} (\bibinfo {year} {2012})},\ \Eprint
  {https://arxiv.org/abs/1201.5108} {arXiv:1201.5108 [nucl-th]} \BibitemShut
  {NoStop}%
\bibitem [{\citenamefont {Tuchin}(2010)}]{Tuchin:2010vs}%
  \BibitemOpen
  \bibfield  {author} {\bibinfo {author} {\bibfnamefont {K.}~\bibnamefont
  {Tuchin}},\ }\bibfield  {title} {\bibinfo {title} {{Synchrotron radiation by
  fast fermions in heavy-ion collisions}},\ }\href
  {https://doi.org/10.1103/PhysRevC.83.039903} {\bibfield  {journal} {\bibinfo
  {journal} {Phys. Rev. C}\ }\textbf {\bibinfo {volume} {82}},\ \bibinfo
  {pages} {034904} (\bibinfo {year} {2010})},\ \bibinfo {note} {[Erratum:
  Phys.Rev.C 83, 039903 (2011)]},\ \Eprint {https://arxiv.org/abs/1006.3051}
  {arXiv:1006.3051 [nucl-th]} \BibitemShut {NoStop}%
\bibitem [{\citenamefont {Tuchin}(2013{\natexlab{a}})}]{Tuchin:2013apa}%
  \BibitemOpen
  \bibfield  {author} {\bibinfo {author} {\bibfnamefont {K.}~\bibnamefont
  {Tuchin}},\ }\bibfield  {title} {\bibinfo {title} {{Time and space dependence
  of the electromagnetic field in relativistic heavy-ion collisions}},\ }\href
  {https://doi.org/10.1103/PhysRevC.88.024911} {\bibfield  {journal} {\bibinfo
  {journal} {Phys. Rev. C}\ }\textbf {\bibinfo {volume} {88}},\ \bibinfo
  {pages} {024911} (\bibinfo {year} {2013}{\natexlab{a}})},\ \Eprint
  {https://arxiv.org/abs/1305.5806} {arXiv:1305.5806 [hep-ph]} \BibitemShut
  {NoStop}%
\bibitem [{\citenamefont {Tuchin}(2013{\natexlab{b}})}]{Tuchin:2013ie}%
  \BibitemOpen
  \bibfield  {author} {\bibinfo {author} {\bibfnamefont {K.}~\bibnamefont
  {Tuchin}},\ }\bibfield  {title} {\bibinfo {title} {{Particle production in
  strong electromagnetic fields in relativistic heavy-ion collisions}},\ }\href
  {https://doi.org/10.1155/2013/490495} {\bibfield  {journal} {\bibinfo
  {journal} {Adv. High Energy Phys.}\ }\textbf {\bibinfo {volume} {2013}},\
  \bibinfo {pages} {490495} (\bibinfo {year} {2013}{\natexlab{b}})},\ \Eprint
  {https://arxiv.org/abs/1301.0099} {arXiv:1301.0099 [hep-ph]} \BibitemShut
  {NoStop}%
\bibitem [{\citenamefont {Zakharov}(2014)}]{Zakharov:2014dia}%
  \BibitemOpen
  \bibfield  {author} {\bibinfo {author} {\bibfnamefont {B.~G.}\ \bibnamefont
  {Zakharov}},\ }\bibfield  {title} {\bibinfo {title} {{Electromagnetic
  response of quark\textendash{}gluon plasma in heavy-ion collisions}},\ }\href
  {https://doi.org/10.1016/j.physletb.2014.08.068} {\bibfield  {journal}
  {\bibinfo  {journal} {Phys. Lett. B}\ }\textbf {\bibinfo {volume} {737}},\
  \bibinfo {pages} {262} (\bibinfo {year} {2014})},\ \Eprint
  {https://arxiv.org/abs/1404.5047} {arXiv:1404.5047 [hep-ph]} \BibitemShut
  {NoStop}%
\bibitem [{\citenamefont {Tuchin}(2016)}]{Tuchin:2015oka}%
  \BibitemOpen
  \bibfield  {author} {\bibinfo {author} {\bibfnamefont {K.}~\bibnamefont
  {Tuchin}},\ }\bibfield  {title} {\bibinfo {title} {{Initial value problem for
  magnetic fields in heavy ion collisions}},\ }\href
  {https://doi.org/10.1103/PhysRevC.93.014905} {\bibfield  {journal} {\bibinfo
  {journal} {Phys. Rev. C}\ }\textbf {\bibinfo {volume} {93}},\ \bibinfo
  {pages} {014905} (\bibinfo {year} {2016})},\ \Eprint
  {https://arxiv.org/abs/1508.06925} {arXiv:1508.06925 [hep-ph]} \BibitemShut
  {NoStop}%
\bibitem [{\citenamefont {Li}\ \emph {et~al.}(2016)\citenamefont {Li},
  \citenamefont {Sheng},\ and\ \citenamefont {Wang}}]{Li:2016tel}%
  \BibitemOpen
  \bibfield  {author} {\bibinfo {author} {\bibfnamefont {H.}~\bibnamefont
  {Li}}, \bibinfo {author} {\bibfnamefont {X.-l.}\ \bibnamefont {Sheng}},\ and\
  \bibinfo {author} {\bibfnamefont {Q.}~\bibnamefont {Wang}},\ }\bibfield
  {title} {\bibinfo {title} {{Electromagnetic fields with electric and chiral
  magnetic conductivities in heavy ion collisions}},\ }\href
  {https://doi.org/10.1103/PhysRevC.94.044903} {\bibfield  {journal} {\bibinfo
  {journal} {Phys. Rev. C}\ }\textbf {\bibinfo {volume} {94}},\ \bibinfo
  {pages} {044903} (\bibinfo {year} {2016})},\ \Eprint
  {https://arxiv.org/abs/1602.02223} {arXiv:1602.02223 [nucl-th]} \BibitemShut
  {NoStop}%
\bibitem [{\citenamefont {Gursoy}\ \emph {et~al.}(2014)\citenamefont {Gursoy},
  \citenamefont {Kharzeev},\ and\ \citenamefont {Rajagopal}}]{Gursoy:2014aka}%
  \BibitemOpen
  \bibfield  {author} {\bibinfo {author} {\bibfnamefont {U.}~\bibnamefont
  {Gursoy}}, \bibinfo {author} {\bibfnamefont {D.}~\bibnamefont {Kharzeev}},\
  and\ \bibinfo {author} {\bibfnamefont {K.}~\bibnamefont {Rajagopal}},\
  }\bibfield  {title} {\bibinfo {title} {{Magnetohydrodynamics, charged
  currents and directed flow in heavy ion collisions}},\ }\href
  {https://doi.org/10.1103/PhysRevC.89.054905} {\bibfield  {journal} {\bibinfo
  {journal} {Phys. Rev. C}\ }\textbf {\bibinfo {volume} {89}},\ \bibinfo
  {pages} {054905} (\bibinfo {year} {2014})},\ \Eprint
  {https://arxiv.org/abs/1401.3805} {arXiv:1401.3805 [hep-ph]} \BibitemShut
  {NoStop}%
\bibitem [{\citenamefont {G\"ursoy}\ \emph {et~al.}(2018)\citenamefont
  {G\"ursoy}, \citenamefont {Kharzeev}, \citenamefont {Marcus}, \citenamefont
  {Rajagopal},\ and\ \citenamefont {Shen}}]{Gursoy:2018yai}%
  \BibitemOpen
  \bibfield  {author} {\bibinfo {author} {\bibfnamefont {U.}~\bibnamefont
  {G\"ursoy}}, \bibinfo {author} {\bibfnamefont {D.}~\bibnamefont {Kharzeev}},
  \bibinfo {author} {\bibfnamefont {E.}~\bibnamefont {Marcus}}, \bibinfo
  {author} {\bibfnamefont {K.}~\bibnamefont {Rajagopal}},\ and\ \bibinfo
  {author} {\bibfnamefont {C.}~\bibnamefont {Shen}},\ }\bibfield  {title}
  {\bibinfo {title} {{Charge-dependent Flow Induced by Magnetic and Electric
  Fields in Heavy Ion Collisions}},\ }\href
  {https://doi.org/10.1103/PhysRevC.98.055201} {\bibfield  {journal} {\bibinfo
  {journal} {Phys. Rev. C}\ }\textbf {\bibinfo {volume} {98}},\ \bibinfo
  {pages} {055201} (\bibinfo {year} {2018})},\ \Eprint
  {https://arxiv.org/abs/1806.05288} {arXiv:1806.05288 [hep-ph]} \BibitemShut
  {NoStop}%
\bibitem [{\citenamefont {Stewart}\ and\ \citenamefont
  {Tuchin}(2018)}]{Stewart:2017zsu}%
  \BibitemOpen
  \bibfield  {author} {\bibinfo {author} {\bibfnamefont {E.}~\bibnamefont
  {Stewart}}\ and\ \bibinfo {author} {\bibfnamefont {K.}~\bibnamefont
  {Tuchin}},\ }\bibfield  {title} {\bibinfo {title} {{Magnetic field in
  expanding quark-gluon plasma}},\ }\href
  {https://doi.org/10.1103/PhysRevC.97.044906} {\bibfield  {journal} {\bibinfo
  {journal} {Phys. Rev. C}\ }\textbf {\bibinfo {volume} {97}},\ \bibinfo
  {pages} {044906} (\bibinfo {year} {2018})},\ \Eprint
  {https://arxiv.org/abs/1710.08793} {arXiv:1710.08793 [nucl-th]} \BibitemShut
  {NoStop}%
\bibitem [{\citenamefont {Csernai}\ \emph {et~al.}(2013)\citenamefont
  {Csernai}, \citenamefont {Magas},\ and\ \citenamefont
  {Wang}}]{Csernai:2013bqa}%
  \BibitemOpen
  \bibfield  {author} {\bibinfo {author} {\bibfnamefont {L.~P.}\ \bibnamefont
  {Csernai}}, \bibinfo {author} {\bibfnamefont {V.~K.}\ \bibnamefont {Magas}},\
  and\ \bibinfo {author} {\bibfnamefont {D.~J.}\ \bibnamefont {Wang}},\
  }\bibfield  {title} {\bibinfo {title} {{Flow Vorticity in Peripheral High
  Energy Heavy Ion Collisions}},\ }\href
  {https://doi.org/10.1103/PhysRevC.87.034906} {\bibfield  {journal} {\bibinfo
  {journal} {Phys. Rev. C}\ }\textbf {\bibinfo {volume} {87}},\ \bibinfo
  {pages} {034906} (\bibinfo {year} {2013})},\ \Eprint
  {https://arxiv.org/abs/1302.5310} {arXiv:1302.5310 [nucl-th]} \BibitemShut
  {NoStop}%
\bibitem [{\citenamefont {Csernai}\ \emph {et~al.}(2014)\citenamefont
  {Csernai}, \citenamefont {Wang}, \citenamefont {Bleicher},\ and\
  \citenamefont {St\"ocker}}]{Csernai:2014ywa}%
  \BibitemOpen
  \bibfield  {author} {\bibinfo {author} {\bibfnamefont {L.~P.}\ \bibnamefont
  {Csernai}}, \bibinfo {author} {\bibfnamefont {D.~J.}\ \bibnamefont {Wang}},
  \bibinfo {author} {\bibfnamefont {M.}~\bibnamefont {Bleicher}},\ and\
  \bibinfo {author} {\bibfnamefont {H.}~\bibnamefont {St\"ocker}},\ }\bibfield
  {title} {\bibinfo {title} {{Vorticity in peripheral collisions at the
  Facility for Antiproton and Ion Research and at the JINR Nuclotron-based Ion
  Collider fAcility}},\ }\href {https://doi.org/10.1103/PhysRevC.90.021904}
  {\bibfield  {journal} {\bibinfo  {journal} {Phys. Rev. C}\ }\textbf {\bibinfo
  {volume} {90}},\ \bibinfo {pages} {021904} (\bibinfo {year}
  {2014})}\BibitemShut {NoStop}%
\bibitem [{\citenamefont {Becattini}\ \emph {et~al.}(2015)\citenamefont
  {Becattini}, \citenamefont {Inghirami}, \citenamefont {Rolando},
  \citenamefont {Beraudo}, \citenamefont {Del~Zanna}, \citenamefont {De~Pace},
  \citenamefont {Nardi}, \citenamefont {Pagliara},\ and\ \citenamefont
  {Chandra}}]{Becattini:2015ska}%
  \BibitemOpen
  \bibfield  {author} {\bibinfo {author} {\bibfnamefont {F.}~\bibnamefont
  {Becattini}}, \bibinfo {author} {\bibfnamefont {G.}~\bibnamefont
  {Inghirami}}, \bibinfo {author} {\bibfnamefont {V.}~\bibnamefont {Rolando}},
  \bibinfo {author} {\bibfnamefont {A.}~\bibnamefont {Beraudo}}, \bibinfo
  {author} {\bibfnamefont {L.}~\bibnamefont {Del~Zanna}}, \bibinfo {author}
  {\bibfnamefont {A.}~\bibnamefont {De~Pace}}, \bibinfo {author} {\bibfnamefont
  {M.}~\bibnamefont {Nardi}}, \bibinfo {author} {\bibfnamefont
  {G.}~\bibnamefont {Pagliara}},\ and\ \bibinfo {author} {\bibfnamefont
  {V.}~\bibnamefont {Chandra}},\ }\bibfield  {title} {\bibinfo {title} {{A
  study of vorticity formation in high energy nuclear collisions}},\ }\href
  {https://doi.org/10.1140/epjc/s10052-015-3624-1} {\bibfield  {journal}
  {\bibinfo  {journal} {Eur. Phys. J. C}\ }\textbf {\bibinfo {volume} {75}},\
  \bibinfo {pages} {406} (\bibinfo {year} {2015})},\ \bibinfo {note} {[Erratum:
  Eur.Phys.J.C 78, 354 (2018)]},\ \Eprint {https://arxiv.org/abs/1501.04468}
  {arXiv:1501.04468 [nucl-th]} \BibitemShut {NoStop}%
\bibitem [{\citenamefont {Deng}\ and\ \citenamefont
  {Huang}(2016)}]{Deng:2016gyh}%
  \BibitemOpen
  \bibfield  {author} {\bibinfo {author} {\bibfnamefont {W.-T.}\ \bibnamefont
  {Deng}}\ and\ \bibinfo {author} {\bibfnamefont {X.-G.}\ \bibnamefont
  {Huang}},\ }\bibfield  {title} {\bibinfo {title} {{Vorticity in Heavy-Ion
  Collisions}},\ }\href {https://doi.org/10.1103/PhysRevC.93.064907} {\bibfield
   {journal} {\bibinfo  {journal} {Phys. Rev. C}\ }\textbf {\bibinfo {volume}
  {93}},\ \bibinfo {pages} {064907} (\bibinfo {year} {2016})},\ \Eprint
  {https://arxiv.org/abs/1603.06117} {arXiv:1603.06117 [nucl-th]} \BibitemShut
  {NoStop}%
\bibitem [{\citenamefont {Jiang}\ \emph {et~al.}(2016)\citenamefont {Jiang},
  \citenamefont {Lin},\ and\ \citenamefont {Liao}}]{Jiang:2016woz}%
  \BibitemOpen
  \bibfield  {author} {\bibinfo {author} {\bibfnamefont {Y.}~\bibnamefont
  {Jiang}}, \bibinfo {author} {\bibfnamefont {Z.-W.}\ \bibnamefont {Lin}},\
  and\ \bibinfo {author} {\bibfnamefont {J.}~\bibnamefont {Liao}},\ }\bibfield
  {title} {\bibinfo {title} {{Rotating quark-gluon plasma in relativistic heavy
  ion collisions}},\ }\href {https://doi.org/10.1103/PhysRevC.94.044910}
  {\bibfield  {journal} {\bibinfo  {journal} {Phys. Rev. C}\ }\textbf {\bibinfo
  {volume} {94}},\ \bibinfo {pages} {044910} (\bibinfo {year} {2016})},\
  \bibinfo {note} {[Erratum: Phys.Rev.C 95, 049904 (2017)]},\ \Eprint
  {https://arxiv.org/abs/1602.06580} {arXiv:1602.06580 [hep-ph]} \BibitemShut
  {NoStop}%
\bibitem [{\citenamefont {Kolomeitsev}\ \emph {et~al.}(2018)\citenamefont
  {Kolomeitsev}, \citenamefont {Toneev},\ and\ \citenamefont
  {Voronyuk}}]{Kolomeitsev:2018svb}%
  \BibitemOpen
  \bibfield  {author} {\bibinfo {author} {\bibfnamefont {E.~E.}\ \bibnamefont
  {Kolomeitsev}}, \bibinfo {author} {\bibfnamefont {V.~D.}\ \bibnamefont
  {Toneev}},\ and\ \bibinfo {author} {\bibfnamefont {V.}~\bibnamefont
  {Voronyuk}},\ }\bibfield  {title} {\bibinfo {title} {{Vorticity and hyperon
  polarization at energies available at JINR Nuclotron-based Ion Collider
  fAcility}},\ }\href {https://doi.org/10.1103/PhysRevC.97.064902} {\bibfield
  {journal} {\bibinfo  {journal} {Phys. Rev. C}\ }\textbf {\bibinfo {volume}
  {97}},\ \bibinfo {pages} {064902} (\bibinfo {year} {2018})},\ \Eprint
  {https://arxiv.org/abs/1801.07610} {arXiv:1801.07610 [nucl-th]} \BibitemShut
  {NoStop}%
\bibitem [{\citenamefont {Deng}\ \emph {et~al.}(2020)\citenamefont {Deng},
  \citenamefont {Huang}, \citenamefont {Ma},\ and\ \citenamefont
  {Zhang}}]{Deng:2020ygd}%
  \BibitemOpen
  \bibfield  {author} {\bibinfo {author} {\bibfnamefont {X.-G.}\ \bibnamefont
  {Deng}}, \bibinfo {author} {\bibfnamefont {X.-G.}\ \bibnamefont {Huang}},
  \bibinfo {author} {\bibfnamefont {Y.-G.}\ \bibnamefont {Ma}},\ and\ \bibinfo
  {author} {\bibfnamefont {S.}~\bibnamefont {Zhang}},\ }\bibfield  {title}
  {\bibinfo {title} {{Vorticity in low-energy heavy-ion collisions}},\ }\href
  {https://doi.org/10.1103/PhysRevC.101.064908} {\bibfield  {journal} {\bibinfo
   {journal} {Phys. Rev. C}\ }\textbf {\bibinfo {volume} {101}},\ \bibinfo
  {pages} {064908} (\bibinfo {year} {2020})},\ \Eprint
  {https://arxiv.org/abs/2001.01371} {arXiv:2001.01371 [nucl-th]} \BibitemShut
  {NoStop}%
\bibitem [{\citenamefont {Xia}\ \emph {et~al.}(2018)\citenamefont {Xia},
  \citenamefont {Li}, \citenamefont {Tang},\ and\ \citenamefont
  {Wang}}]{Xia:2018tes}%
  \BibitemOpen
  \bibfield  {author} {\bibinfo {author} {\bibfnamefont {X.-L.}\ \bibnamefont
  {Xia}}, \bibinfo {author} {\bibfnamefont {H.}~\bibnamefont {Li}}, \bibinfo
  {author} {\bibfnamefont {Z.-B.}\ \bibnamefont {Tang}},\ and\ \bibinfo
  {author} {\bibfnamefont {Q.}~\bibnamefont {Wang}},\ }\bibfield  {title}
  {\bibinfo {title} {{Probing vorticity structure in heavy-ion collisions by
  local $\Lambda$ polarization}},\ }\href
  {https://doi.org/10.1103/PhysRevC.98.024905} {\bibfield  {journal} {\bibinfo
  {journal} {Phys. Rev. C}\ }\textbf {\bibinfo {volume} {98}},\ \bibinfo
  {pages} {024905} (\bibinfo {year} {2018})},\ \Eprint
  {https://arxiv.org/abs/1803.00867} {arXiv:1803.00867 [nucl-th]} \BibitemShut
  {NoStop}%
\bibitem [{\citenamefont {Velikhov}(1959)}]{velikhov1959stability}%
  \BibitemOpen
  \bibfield  {author} {\bibinfo {author} {\bibfnamefont {E.}~\bibnamefont
  {Velikhov}},\ }\bibfield  {title} {\bibinfo {title} {Stability of an ideally
  conducting liquid flowing between cylinders rotating in a magnetic field},\
  }\href@noop {} {\bibfield  {journal} {\bibinfo  {journal} {Sov. Phys. JETP}\
  }\textbf {\bibinfo {volume} {36}},\ \bibinfo {pages} {995} (\bibinfo {year}
  {1959})}\BibitemShut {NoStop}%
\bibitem [{\citenamefont {Chandrasekhar}(1960)}]{chandrasekhar1960stability}%
  \BibitemOpen
  \bibfield  {author} {\bibinfo {author} {\bibfnamefont {S.}~\bibnamefont
  {Chandrasekhar}},\ }\bibfield  {title} {\bibinfo {title} {The stability of
  non-dissipative couette flow in hydromagnetics},\ }\href@noop {} {\bibfield
  {journal} {\bibinfo  {journal} {Proceedings of the National Academy of
  Sciences}\ }\textbf {\bibinfo {volume} {46}},\ \bibinfo {pages} {253}
  (\bibinfo {year} {1960})}\BibitemShut {NoStop}%
\bibitem [{\citenamefont {Balbus}\ and\ \citenamefont
  {Hawley}(1991)}]{Balbus:1991ay}%
  \BibitemOpen
  \bibfield  {author} {\bibinfo {author} {\bibfnamefont {S.~A.}\ \bibnamefont
  {Balbus}}\ and\ \bibinfo {author} {\bibfnamefont {J.~F.}\ \bibnamefont
  {Hawley}},\ }\bibfield  {title} {\bibinfo {title} {{A powerful local shear
  instability in weakly magnetized disks. 1. Linear analysis. 2. Nonlinear
  evolution}},\ }\href {https://doi.org/10.1086/170270} {\bibfield  {journal}
  {\bibinfo  {journal} {Astrophys. J.}\ }\textbf {\bibinfo {volume} {376}},\
  \bibinfo {pages} {214} (\bibinfo {year} {1991})}\BibitemShut {NoStop}%
\bibitem [{\citenamefont {Wang}\ and\ \citenamefont
  {Huang}(2024)}]{Wang:2023imu}%
  \BibitemOpen
  \bibfield  {author} {\bibinfo {author} {\bibfnamefont {S.}~\bibnamefont
  {Wang}}\ and\ \bibinfo {author} {\bibfnamefont {X.-G.}\ \bibnamefont
  {Huang}},\ }\bibfield  {title} {\bibinfo {title} {{Chiral magnetovortical
  instability}},\ }\href {https://doi.org/10.1103/PhysRevD.109.L121302}
  {\bibfield  {journal} {\bibinfo  {journal} {Phys. Rev. D}\ }\textbf {\bibinfo
  {volume} {109}},\ \bibinfo {pages} {L121302} (\bibinfo {year} {2024})},\
  \Eprint {https://arxiv.org/abs/2307.06746} {arXiv:2307.06746 [nucl-th]}
  \BibitemShut {NoStop}%
\bibitem [{\citenamefont {Aarts}\ \emph {et~al.}(2007)\citenamefont {Aarts},
  \citenamefont {Allton}, \citenamefont {Foley}, \citenamefont {Hands},\ and\
  \citenamefont {Kim}}]{Aarts:2007wj}%
  \BibitemOpen
  \bibfield  {author} {\bibinfo {author} {\bibfnamefont {G.}~\bibnamefont
  {Aarts}}, \bibinfo {author} {\bibfnamefont {C.}~\bibnamefont {Allton}},
  \bibinfo {author} {\bibfnamefont {J.}~\bibnamefont {Foley}}, \bibinfo
  {author} {\bibfnamefont {S.}~\bibnamefont {Hands}},\ and\ \bibinfo {author}
  {\bibfnamefont {S.}~\bibnamefont {Kim}},\ }\bibfield  {title} {\bibinfo
  {title} {{Spectral functions at small energies and the electrical
  conductivity in hot, quenched lattice QCD}},\ }\href
  {https://doi.org/10.1103/PhysRevLett.99.022002} {\bibfield  {journal}
  {\bibinfo  {journal} {Phys. Rev. Lett.}\ }\textbf {\bibinfo {volume} {99}},\
  \bibinfo {pages} {022002} (\bibinfo {year} {2007})},\ \Eprint
  {https://arxiv.org/abs/hep-lat/0703008} {arXiv:hep-lat/0703008} \BibitemShut
  {NoStop}%
\bibitem [{\citenamefont {Ding}\ \emph {et~al.}(2011)\citenamefont {Ding},
  \citenamefont {Francis}, \citenamefont {Kaczmarek}, \citenamefont {Karsch},
  \citenamefont {Laermann},\ and\ \citenamefont {Soeldner}}]{Ding:2010ga}%
  \BibitemOpen
  \bibfield  {author} {\bibinfo {author} {\bibfnamefont {H.~T.}\ \bibnamefont
  {Ding}}, \bibinfo {author} {\bibfnamefont {A.}~\bibnamefont {Francis}},
  \bibinfo {author} {\bibfnamefont {O.}~\bibnamefont {Kaczmarek}}, \bibinfo
  {author} {\bibfnamefont {F.}~\bibnamefont {Karsch}}, \bibinfo {author}
  {\bibfnamefont {E.}~\bibnamefont {Laermann}},\ and\ \bibinfo {author}
  {\bibfnamefont {W.}~\bibnamefont {Soeldner}},\ }\bibfield  {title} {\bibinfo
  {title} {{Thermal dilepton rate and electrical conductivity: An analysis of
  vector current correlation functions in quenched lattice QCD}},\ }\href
  {https://doi.org/10.1103/PhysRevD.83.034504} {\bibfield  {journal} {\bibinfo
  {journal} {Phys. Rev. D}\ }\textbf {\bibinfo {volume} {83}},\ \bibinfo
  {pages} {034504} (\bibinfo {year} {2011})},\ \Eprint
  {https://arxiv.org/abs/1012.4963} {arXiv:1012.4963 [hep-lat]} \BibitemShut
  {NoStop}%
\bibitem [{\citenamefont {Amato}\ \emph {et~al.}(2014)\citenamefont {Amato},
  \citenamefont {Aarts}, \citenamefont {Allton}, \citenamefont {Giudice},
  \citenamefont {Hands},\ and\ \citenamefont {Skullerud}}]{Amato:2013oja}%
  \BibitemOpen
  \bibfield  {author} {\bibinfo {author} {\bibfnamefont {A.}~\bibnamefont
  {Amato}}, \bibinfo {author} {\bibfnamefont {G.}~\bibnamefont {Aarts}},
  \bibinfo {author} {\bibfnamefont {C.}~\bibnamefont {Allton}}, \bibinfo
  {author} {\bibfnamefont {P.}~\bibnamefont {Giudice}}, \bibinfo {author}
  {\bibfnamefont {S.}~\bibnamefont {Hands}},\ and\ \bibinfo {author}
  {\bibfnamefont {J.-I.}\ \bibnamefont {Skullerud}},\ }\bibfield  {title}
  {\bibinfo {title} {{Transport coefficients of the QGP}},\ }\href
  {https://doi.org/10.22323/1.187.0176} {\bibfield  {journal} {\bibinfo
  {journal} {PoS}\ }\textbf {\bibinfo {volume} {LATTICE2013}},\ \bibinfo
  {pages} {176} (\bibinfo {year} {2014})},\ \Eprint
  {https://arxiv.org/abs/1310.7466} {arXiv:1310.7466 [hep-lat]} \BibitemShut
  {NoStop}%
\bibitem [{\citenamefont {Buzzegoli}\ and\ \citenamefont
  {Tuchin}(2023)}]{Buzzegoli:2023yut}%
  \BibitemOpen
  \bibfield  {author} {\bibinfo {author} {\bibfnamefont {M.}~\bibnamefont
  {Buzzegoli}}\ and\ \bibinfo {author} {\bibfnamefont {K.}~\bibnamefont
  {Tuchin}},\ }\bibfield  {title} {\bibinfo {title} {{Electromagnetic radiation
  at extreme angular velocity}},\ }\href
  {https://doi.org/10.1007/JHEP12(2023)113} {\bibfield  {journal} {\bibinfo
  {journal} {JHEP}\ }\textbf {\bibinfo {volume} {12}},\ \bibinfo {pages}
  {113}},\ \Eprint {https://arxiv.org/abs/2308.10349} {arXiv:2308.10349
  [hep-ph]} \BibitemShut {NoStop}%
\bibitem [{\citenamefont {Zel'Dovich}(1971)}]{zel1971generation}%
  \BibitemOpen
  \bibfield  {author} {\bibinfo {author} {\bibfnamefont {Y.~B.}\ \bibnamefont
  {Zel'Dovich}},\ }\bibfield  {title} {\bibinfo {title} {Generation of waves by
  a rotating body},\ }\href@noop {} {\bibfield  {journal} {\bibinfo  {journal}
  {Soviet Journal of Experimental and Theoretical Physics Letters}\ }\textbf
  {\bibinfo {volume} {14}},\ \bibinfo {pages} {180} (\bibinfo {year}
  {1971})}\BibitemShut {NoStop}%
\bibitem [{\citenamefont {ZEL'DOVICH}(1972)}]{zel1972amplification}%
  \BibitemOpen
  \bibfield  {author} {\bibinfo {author} {\bibfnamefont {I.}~\bibnamefont
  {ZEL'DOVICH}},\ }\bibfield  {title} {\bibinfo {title} {Amplification of
  cylindrical electromagnetic waves reflected from a rotating body},\
  }\href@noop {} {\bibfield  {journal} {\bibinfo  {journal} {Soviet
  Physics-JETP}\ }\textbf {\bibinfo {volume} {35}},\ \bibinfo {pages} {1085}
  (\bibinfo {year} {1972})}\BibitemShut {NoStop}%
\bibitem [{\citenamefont {Abramowitz}\ and\ \citenamefont
  {Stegun}(1965)}]{abramowitz1965handbook}%
  \BibitemOpen
  \bibfield  {author} {\bibinfo {author} {\bibfnamefont {M.}~\bibnamefont
  {Abramowitz}}\ and\ \bibinfo {author} {\bibfnamefont {I.}~\bibnamefont
  {Stegun}},\ }\href {https://books.google.com/books?id=MtU8uP7XMvoC} {\emph
  {\bibinfo {title} {Handbook of Mathematical Functions: With Formulas, Graphs,
  and Mathematical Tables}}},\ Applied mathematics series\ (\bibinfo
  {publisher} {Dover Publications},\ \bibinfo {year} {1965})\BibitemShut
  {NoStop}%
\bibitem [{\citenamefont {Zwillinger}(2014)}]{zwillinger2014table}%
  \BibitemOpen
  \bibfield  {author} {\bibinfo {author} {\bibfnamefont {D.}~\bibnamefont
  {Zwillinger}},\ }\href {https://books.google.com/books?id=NjnLAwAAQBAJ}
  {\emph {\bibinfo {title} {Table of Integrals, Series, and Products}}}\
  (\bibinfo  {publisher} {Academic Press},\ \bibinfo {year} {2014})\BibitemShut
  {NoStop}%
\end{thebibliography}%

\appendix

\section{Calculation of $\mathcal{B}_m(0,k_{\perp},k_z)$}\label{sec:Coefficients}
In this section, we perform the integration in Eq.~\eqref{b3.1}, to find the amplitudes $\mathcal{B}_{m}(t,k_{\perp},k_z)$. We first perform the $\phi$-integration followed by the integration over $\rho$. The integral over $z$ can be performed separately resulting in
\begin{align}
    \int\limits_{-\infty}^{\infty}dz\, e^{-ik_z z}e^{-\frac{z^2}{\Delta_z^2}} = \sqrt{\pi}\Delta_z\exp\left(-\frac{k_z^2}{4}\Delta_z^2\right)\,.
\end{align}
For the $\phi$-integral we write 
\begin{align}
    \rho^2\left(\frac{\cos^2\phi}{\Delta_x^2}+\frac{\sin^2\phi}{\Delta_y^2}\right) = \frac{\rho^2}{2}\left[\frac{1}{\Delta_+} + \frac{1}{\Delta_-}\cos(2\phi)\right]\,.
\end{align}
This leads to 
\begin{align}
    &\mathcal{B}_m(0,k_{\perp},k_z) \nonumber \\
    &=B_0\bm{\hat{z}}\sqrt{\pi}\Delta_z e^{-\frac{1}{4}k_z^2\Delta_z^2}\int\limits_{0}^{\infty}d\rho\ \rho J_{m}(k_{\perp}\rho)e^{-\frac{\rho^2}{2\Delta_+}}\int\limits_{0}^{2\pi}\frac{d\phi}{2\pi}e^{-im\phi}e^{-\frac{\rho^2}{2\Delta_-}\cos(2\phi)}
\end{align}
In the last line we end up with $\phi$ integral of the form 
\begin{align}
    \mathcal{I}_n(\alpha) = \int\limits_{0}^{2\pi}\frac{d\phi}{2\pi}e^{-in\phi}e^{-\alpha\cos(2\phi)}\ ,
\end{align}
where $\alpha = \rho^2 / (2\Delta_{-})$.
To do this integration, we make a change of variable $\phi^{\prime} = 2 (\phi - \pi)$. Then
\begin{align}
    \mathcal{I}_m(\alpha) &= \frac{1}{2}\int\limits_{-2\pi}^{2\pi}\frac{d\phi^{\prime}}{2\pi}e^{-im\left(\frac{\phi^{\prime}}{2}+\pi\right)}e^{-\alpha\cos(\phi^{\prime}+2\pi)} \nonumber \\
    &= \frac{(-1)^m}{2}\int\limits_{-2\pi}^{2\pi}\frac{d\phi^{\prime}}{2\pi}e^{-im\frac{\phi^{\prime}}{2}}e^{-\alpha\cos\phi^{\prime}}\nonumber \\
    &= \frac{(-1)^m}{2}\int\limits_{-2\pi}^{2\pi}\frac{d\phi^{\prime}}{2\pi}\left[\cos\left(m\frac{\phi^{\prime}}{2}\right) -i \sin\left(m\frac{\phi^{\prime}}{2}\right)\right]e^{-\alpha\cos\phi^{\prime}}
\end{align}
The integral over sine vanishes, while the finite part can be expressed in terms of the following function
\begin{align}
    \mathcal{I}_m(\alpha) &= (-1)^m \int\limits_{0}^{2\pi}\frac{d\phi^{\prime}}{2\pi}\ \cos\left(\frac{m\phi^{\prime}}{2}\right)e^{-\alpha \cos\phi^{\prime}}
\end{align}
Now, another change of variable $\psi = \phi^{\prime}-\pi$ yields
\begin{align}
    \mathcal{I}_m(\alpha) &= (-1)^m \int\limits_{-\pi}^{\pi}\frac{d\psi}{2\pi}\cos\left(\frac{m\psi}{2}+\frac{m\pi}{2}\right)e^{-\alpha\cos(\psi+\pi)} \nonumber \\
    &=(-1)^m\int\limits_{-\pi}^{\pi}\frac{d\psi}{2\pi}\left[\cos\left(\frac{m\psi}{2}\right)\cos\left(\frac{m\pi}{2}\right)-\sin\left(\frac{m\psi}{2}\right)\sin\left(\frac{m\pi}{2}\right)\right]e^{\alpha\cos\psi}\nonumber \\
    &=(-1)^m\cos\left(\frac{m\pi}{2}\right)\int\limits_{0}^{\pi}\frac{d\psi}{\pi}\cos\left(\frac{m\psi}{2}\right)e^{\alpha\cos\psi}
\end{align}
In the last line the integrand involving sine is odd in $\psi$. However, when $m$ is odd, the integration vanishes due to the cosine term. So we do not need to worry about the integral when $m$ is odd. For the even $m$, we have $m/2\in \mathbbm{Z}$. In this case, we can use the following representation of modified Bessel function from Eq.~(9.6.19) of \cite{abramowitz1965handbook} which read as
\begin{align}
    I_{\nu}(z) = \frac{1}{\pi}\int_{0}^{\pi}d\theta\,e^{z\cos\theta} \cos(\nu\theta), \quad \nu\in\mathbbm{Z}
\end{align}
Therefore, we arrive at 
\begin{align}
    \mathcal{I}_{m}(\alpha) = \begin{cases}
        \cos\left(m\frac{\pi}{2}\right)I_{\frac{m}{2}}(\alpha),& m\text{ even} \\
        0,& m\text{ odd}
    \end{cases}
\end{align}
Thus we get after the $\phi$-integration,
\begin{align}
    &\mathcal{B}_{m}(0,k_{\perp},k_z)  \nonumber \\
    &= B_0 \bm{\hat{z}} \sqrt{\pi}\Delta_ze^{-\frac{1}{4}k_z^2\Delta_z^2}\cos\left(\frac{m\pi}{2}\right)\int\limits_{0}^{\infty}d\rho\ \rho\,e^{-\rho^2/(2\Delta_+)}J_{m}(k_{\perp}\rho) I_{\frac{m}{2}}\left(\frac{\rho^2}{2\Delta_{-}}\right),\quad m\,\,\text{ even} 
\end{align}
and $\mathcal{B}_{m}(0,k_{\perp},k_z) = 0$ when $m$ is odd. To perform the $\rho$ integration, we exploit the identity 
\begin{align}
    I_{\nu}(z) = i^{-\nu}J_{\nu}(z), \quad\text{for }\nu\in \mathbbm{Z}, \label{eq:app:BesselI_def}
\end{align} 
to write down
\begin{align}
    &\mathcal{B}_{m}(0,k_{\perp},k_z)  \nonumber \\
    &= B_0 \bm{\hat{z}} \sqrt{\pi}\Delta_ze^{-\frac{1}{4}k_z^2\Delta_z^2}\cos\left(\frac{m\pi}{2}\right)i^{-m/2}\int\limits_{0}^{\infty}d\rho\ \rho\,e^{-\rho^2/(2\Delta_+)}J_{m}(k_{\perp}\rho) J_{\frac{m}{2}}\left(i\frac{\rho^2}{2\Delta_{-}}\right),\quad m\,\,\text{ even} 
\end{align}
Next we can use the identity 6.651(6) of~\cite{zwillinger2014table}
\begin{align}
    &\int\limits_{0}^{\infty} dx\ xe^{-\frac{1}{4}\alpha x^2}J_{\nu}(cx)J_{\nu/2}\left(\frac{1}{4}bx^2\right)  \nonumber \\
    &=\frac{2}{\sqrt{\alpha^2+b^2}}\exp\left(-\frac{\alpha c^2}{\alpha^2+b^2}\right)J_{\nu/2}\left(\frac{bc^2}{\alpha^2+b^2}\right), \quad c>0,\ \text{Re}(\alpha)>|\text{Im}(b)|,\ \text{Re}(\nu) > 1
\end{align}
to compute the $\rho$ integration. \\
After identifying $\alpha$, $b$ and $c$ as $2/\Delta_+$, $2i/\Delta_-$ and $k_{\perp}$, respectively, we write
\begin{align}
    &\mathcal{B}_{m}(0,k_{\perp},k_z)  \nonumber \\
    &= B_0 \bm{\hat{z}} \sqrt{\pi}\Delta_ze^{-\frac{1}{4}k_z^2\Delta_z^2}\cos\left(\frac{m\pi}{2}\right)i^{-m/2}\frac{\Delta_x\Delta_y}{2}\exp\left(-\frac{k_{\perp}^2}{8}(\Delta_x^2+\Delta_y^2)\right)J_{\frac{m}{2}}\left(i\frac{k_{\perp}^2}{8}(\Delta_y^2-\Delta_x^2)\right),\quad m\,\,\text{ even} 
\end{align}
Reusing the identity in Eq.~\eqref{eq:app:BesselI_def} and $I_{\nu}(z) = (-1)^{\nu}I_{\nu}(-z)$, we finally arrive at 
\begin{align}
    &\mathcal{B}_{m}(0,k_{\perp},k_z)  \nonumber \\
    &= B_0 \bm{\hat{z}} \sqrt{\pi}\frac{\Delta_x\Delta_y\Delta_z}{2}e^{-\frac{1}{4}k_z^2\Delta_z^2}\cos\left(\frac{m\pi}{2}\right)(-1)^{m/2}\exp\left(-\frac{k_{\perp}^2}{8}(\Delta_x^2+\Delta_y^2)\right)I_{\frac{m}{2}}\left(\frac{k_{\perp}^2}{8}(\Delta_x^2-\Delta_y^2)\right),\quad m\,\,\text{ even}
\end{align}


\section{Stationary plasma: $\Omega=0$.}\label{sec:app:Omega=0}
In this section, we compute $\bm{B}(t,\bm{r})$ for $\Omega = 0$ by solving 
\begin{align}
    \left(\frac{\partial^2}{\partial t^2}-\nabla^2 + \sigma\frac{\partial}{\partial t}\right)\bm{B}(t,\bm{r}) = \bm{0}\,.
\end{align}
with the initial condition as given by Eq.~\eqref{b3.1}. Writing $\bm{B}$ in Fourier basis
\begin{align}
    \bm{B}(t,\bm{r}) = \int\!\frac{d^3k}{(2\pi)^3}e^{i\bm{k}.\bm{r}}\bm{\mathcal{B}}(t,\bm{k}), \label{eq:app:B_Omega0}
\end{align}
we get 
\begin{align}
    \partial_t^2\bm{\mathcal{B}}(t,\bm{k}) + \sigma\partial_t\bm{\mathcal{B}}(t,\bm{k}) + k^2\bm{\mathcal{B}}(t,\bm{k}) = 0.
\end{align}
The last equation is the same as Eq.~\eqref{a9} except with $\Omega = 0$. Therefore, we can write the time dependence of Fourier coefficients
\begin{align}
    \bm{\mathcal{B}}(t,\bm{k}) &= e^{-\frac{\sigma t}{2}}\left(\left[\cos\left(\sqrt{k^2-\frac{\sigma^2}{4}}\ t\right)+\frac{\sigma}{2\sqrt{k^2-\frac{\sigma^2}{4}}}\sin\left(\sqrt{k^2-\frac{\sigma^2}{4}}\ t\right)\right]\bm{\mathcal{B}}(0,\bm{k})\right.\nonumber \\ 
    &\hspace{7cm}\left.+ \frac{1}{\sqrt{k^2-\frac{\sigma^2}{4}}}\sin\left(\sqrt{k^2-\frac{\sigma^2}{4}}\ t\right)\bm{\dot{\mathcal{B}}}(0,\bm{k})\right),
\end{align}
where 
\begin{align}
    \bm{\mathcal{B}}(0,\bm{k}) = \int d^3r\ e^{-i\bm{k}.\bm{r}}\bm{B}(0,\bm{r}), \\
    \bm{\dot{\mathcal{B}}}(0,\bm{k}) = \int d^3r\ e^{-i\bm{k}.\bm{r}}\dot{\bm{B}}(0,\bm{r})
\end{align}
In our model, we have 
\begin{align}
    \mathcal{B}_z(0,\bm{k}) = B_0\pi^{3/2}\Delta_x\Delta_y\Delta_z\exp\left[-\frac{1}{4}\left(k_x^2\Delta_x^2 + k_y^2\Delta_y^2 + k_z^2\Delta_z^2\right)\right],\quad \dot{\mathcal{B}}_z(0,\bm{k}) = 0. \label{eq:app:mathcalBk_Omega0}
\end{align}
In our model, we have taken $\Delta_x = \Delta_z = \Delta$, $\Delta_y = \Delta/\gamma$. Substituting Eq.~\eqref{eq:app:mathcalBk_Omega0} into Eq.~\eqref{eq:app:B_Omega0}, we get 
\begin{align}
    &B_z(t,\bm{k}) = B_0\pi^{3/2}\Delta^3e^{-\frac{\sigma t}{2}}\int\!\frac{d^3k}{(2\pi)^3} e^{i\bm{k}.\bm{r}}\left(\left[\cos\sqrt{k^2-\frac{\sigma^2}{4}}\,t + \frac{\sigma}{2\sqrt{k^2-\frac{\sigma^2}{4}}}\sin\sqrt{k^2-\frac{\sigma^2}{4}}\,t\right]\right)\nonumber \\
    &\hspace{10cm}\times\exp\left[-\frac{\Delta^2}{4}\left(k_y^2+\frac{k_T^2}{\gamma^2}\right)\right],
\end{align}
where $k_T=\sqrt{k_x^2+k_z^2}$. The integrand exhibits rotational symmetry on the plane transverse to the $y$ direction. So, we can orient our $k_x$-axis along $\bm{\rho}_T$, where $\bm{r} = (\bm{\rho}_T, y)$. It allows us to write the measure $d^3k$ as $k_Tdk_Td\psi_kdk_y$, with $k_x = k_T\cos\psi_k$, $k_z=k_T\sin\psi_k$, which gives
\begin{align}
    &B_z(t,\bm{k}) = \frac{B_0\pi^{3/2}\Delta^3}{8\pi^3}e^{-\frac{\sigma t}{2}}\int_{0}^{\infty}dk_T\, k_T\int_{-\infty}^{\infty}dk_y\int_{0}^{2\pi}d\psi_k\ e^{i(k_T\rho_T\cos\psi_k + k_yy)}g(t,k_{T},k_y) \nonumber \\
    &= \frac{B_0\pi^{3/2}\Delta^3}{4\pi^3}e^{-\frac{\sigma t}{2}}\int_{-\infty}^{0}dk_y\cos (k_yy)\int_{0}^{\infty}dk_T\ k_T\,g(t,k_T,k_y)\int_{0}^{2\pi}d\psi_k\ e^{ik_T\rho_T\cos\psi_k},
\end{align}
where 
\begin{align}
    &g(t,k_T,k_y) = \nonumber \\
    &\left(\left[\cos\sqrt{k_T^2+k_y^2-\frac{\sigma^2}{4}}\,t + \frac{\sigma}{2\sqrt{k_T^2+k_y^2-\frac{\sigma^2}{4}}}\sin\sqrt{k_T^2+k_y^2-\frac{\sigma^2}{4}}\,t\right]\right)\exp\left[-\frac{\Delta^2}{4}\left(k_y^2+\frac{k_T^2}{\gamma^2}\right)\right] \label{eq:app:g(t,k_T,k_y)}
\end{align}
From the identity 
\begin{align}
    2\pi J_0(k_T\rho_T) = \int_{0}^{2\pi}d\psi_k\ e^{ik_T\rho_T\cos\psi_k},
\end{align}
we have the simplified form of $B_z$ at $\Omega = 0$ as
\begin{align}
    B_z(t,\bm{r}) = \frac{B_0\pi^{3/2}\Delta^3}{2\pi^2}e^{-\frac{\sigma t}{2}}\int^{\infty}_{0}dk_y\cos (k_yy)\int_{0}^{\infty}dk_T\ k_T\,J_0(k_T\rho_T)\,g(t,k_T,k_y).
\end{align}
where $g(t,k_T,k_y)$ is given by Eq.~\eqref{eq:app:g(t,k_T,k_y)}.
\end{document}